\documentclass{aastex}
\usepackage{emulateapj5}

\newcommand{\RCW}{{PKS~J0859-4731}}
\newcommand{\carneb}{{the Carina Nebula}}
\newcommand{\paperh}{{Paper II}}
\newcommand{\paperp}{{Paper III}}

\begin{document}

\slugcomment{Published in the Astrophysical Journal, 568, 28}

\title{Experiment Design and First Season Observations with the Degree
Angular Scale Interferometer}

\author{E.\ M.\ Leitch, C.\ Pryke, N.\ W.\ Halverson, J.\ Kovac,
G.\ Davidson, S.\ LaRoque, E.\ Schartman\altaffilmark{1}, J.\ Yamasaki, J.\ E.\ Carlstrom}
\affil{University of Chicago, 5640 South Ellis Ave., Chicago, IL 60637}
\author{W.\ L.\ Holzapfel}
\affil{University of California, 426 Le Conte Hall, Berkeley,
CA 94720}
\author{M.\ Dragovan}
\affil{Jet Propulsion Laboratory,
California Institute of Technology,
4800 Oak Grove Drive,
Pasadena, CA 91109}
\author{J.\ K.\ Cartwright, B.\ S.\ Mason, S.\ Padin,
T.\ J.\ Pearson, A.\ C.\ S.\ Readhead, M.\ C.\ Shepherd}
\affil{California Institute of Technology,
1200 East California Boulevard,
Pasadena, CA 91125}
\altaffiltext{1}{Currently at Princeton University}

\begin{abstract}

We describe the instrumentation, experiment design and data reduction
for the first season of observations with the Degree Angular Scale
Interferometer (DASI), a compact microwave interferometer designed to
measure anisotropy of the Cosmic Microwave Background (CMB) on degree
and sub-degree scales ($l \simeq $~100--900).  The telescope was deployed at
the Amundsen-Scott South Pole research station during the 1999--2000
austral summer and conducted observations of the CMB throughout the
following austral winter.  In its first season of observations, DASI has
mapped CMB fluctuations in 32 fields, each $3\fdg4$ across, with high
sensitivity.
\end{abstract}

\keywords{cosmology: cosmic microwave background---cosmology: observations
	---techniques:interferometric}

\section{Introduction}

The use of the CMB angular power spectrum to constrain cosmological
parameters has been the subject of much recent literature \markcite{hu96a}(see,
e.g., Hu \& White 1996,  for a discussion of salient features of the power
spectrum).  We are at a point where the ability to resolve fine-scale
structure in the power spectrum is no longer limited primarily by
detector sensitivity, but by experiment design, understanding of
calibration uncertainties and careful control of systematics.  The
comparison of complementary measurements by experiments of radically
different design will prove critical to an understanding of what
exactly the CMB is telling us.

Because they directly sample Fourier components of the sky,
interferometers are uniquely suited to measurement of the CMB power
spectrum and offer a completely independent technique, free of many of
the systematics which must be carefully controlled in scanning
experiments.  The Degree Angular Scale Interferometer (DASI), along
with its companion instrument the CBI \markcite{pearson00}({Pearson} {et~al.} 2000), and the VSA
\markcite{jones96}({Jones} 1996), is one of a new generation of ultra-compact microwave
interferometers designed to measure anisotropy in the CMB.  With 13
elements operating in ten 1-GHz bands from 26--36~GHz, DASI provides
dense sampling of the power spectrum from $l = 100$ to $l = 900$,
angular scales spanning the first three harmonically related acoustic
peaks in $\Omega\sim 1$ cosmologies.

This paper describes details of the instrument, design of the
experiment, and calibration and reduction of the Fourier-plane
visibility data.  Analysis of these data and extraction of the angular
power spectrum are presented in \markcite{halverson01}Halverson {et~al.} (2001,  hereafter \paperh), 
while limits on cosmological parameters from the DASI data
are given in \markcite{pryke01}Pryke {et~al.} (2001,  hereafter \paperp).

In sections \S\ref{sec:instrument} and \S\ref{sec:interferometer} of
this paper, we discuss details of the instrument design.  A
description of the site is given in \S\ref{sec:site}.  We discuss
foregrounds in \S\ref{sec:foregrounds}.  The CMB observations are
described in \S\ref{sec:obs} and
\S\ref{sec:calibration}.  Data reduction and analysis are presented in
\S\ref{sec:reduc} and \S\ref{sec:ptsrc}, and results are presented in
\S\ref{sec:results}.

\section{Instrument}
\label{sec:instrument}

\begin{figure*}[th]
\plotone{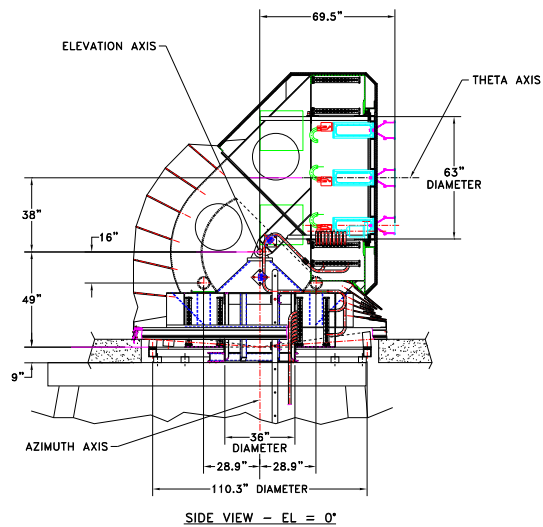}
\caption{Schematic of the DASI telescope.  Shown is a side view of the
instrument, at EL = 0, with the faceplate pointing to the right.}
\label{fig:schematic}
\epsscale{1.0}
\end{figure*}

\subsection{Telescope Mount}

The telescope is situated atop the inner of two concentric towers
attached to the Martin A. Pomerantz Observatory (MAPO), 0.7~km from
the geographic South Pole.  The inner tower is mechanically isolated
from the outer tower to minimize vibrations transmitted from the
building to the telescope.  Although diurnal variations in the ambient
temperature at the Pole are extremely small (see \S\ref{sec:site}),
the legs of the inner tower are insulated to minimize tilt of
the azimuth ring from differential thermal expansion by direct solar
heating during the summer months.

A room beneath the telescope, attached to the outer tower, houses
helium compressors, drive amplifiers, and an air handling unit for
managing waste heat from the telescope and compressors.  The interior
of the telescope opens directly onto the compressor room, providing
access to drive systems, receivers and electronics even in mid-winter,
when the darkness and extreme cold (as low as $-80\degr$~C ambient)
severely restrict outside activity.  An insulated fabric bellows
permits motion in the elevation axis while maintaining the interior of
the telescope and drive assemblies at room temperature using only
waste heat from the telescope systems.
Tested for flexibility to $-100\degr$~C, the bellows has thus far
functioned perfectly at polar temperatures.

The telescope mount is an altitude-azimuth design, employing a
counterbalanced gear and pinion elevation drive, for tracking and
pointing stability.  Heavy box steel construction lends the mount
extreme rigidity and immunity to flexure; the combined weight of the
telescope, when fully equipped and operational, is approximately
35,000~lbs.

The absolute pointing of the telescope is derived from observations of
bright stars made with a small optical camera mounted on the
faceplate.  In an automated procedure, approximately 80 positions for
stars distributed throughout the sky are acquired using a frame
grabber.  These observations are fit with an eight parameter pointing
model, yielding $\sim20''$ rms residuals.  Limits on the deviation of
the radio pointing from this model are discussed in \S\ref{sec:ptsrc}.

\subsection{Faceplate}

\begin{figure*}[th]
\plotone{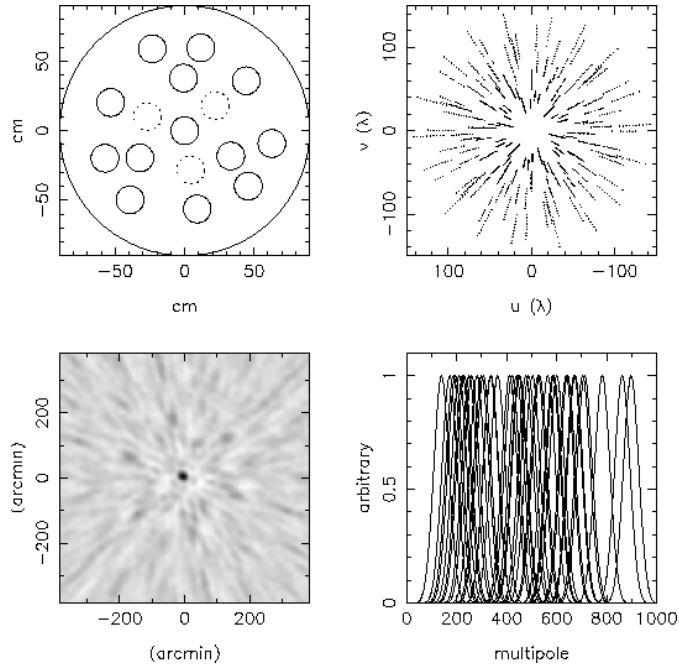}
\caption{Top left: The DASI faceplate configuration.  Solid circles
represent locations of the 13 antennas.  (Dashed circles are additional
faceplate slots.)  Top right: The resulting instantaneous Fourier
plane coverage of the interferometer, showing radial extension due to
the frequency dependence of the $(u,v)$ coordinates. Bottom left: one
of the brightest point sources in the DASI fields, showing the
synthesized beam pattern.  Radial features in the image are artifacts
of the sampling.  Bottom right: The effective $l$-space window
functions of the 26 distinct DASI baselines, shown for 26.5 and 35.5
GHz.  Note that this represents only 1/5 of DASI's instantaneous
$l$-space sampling.}
\label{fig:int}
\end{figure*}

The interferometer has 13 primary antenna elements, arranged in a
three-fold symmetric pattern on a rigid faceplate attached to the
elevation cradle.  The locations of the antennas in the faceplate were
numerically optimized to provide nearly uniform sampling over the
multipole range probed by DASI, $l\sim 100$--900 (see
Figure~\ref{fig:int}).  The rigid faceplate greatly simplifies the
design of the correlator, since unlike conventional tracking arrays,
projected baseline lengths for a co-planar array are independent of
the pointing center, and tracking delays are not required.

The faceplate can also be rotated about its axis.  In combination with
the three-fold symmetry, this feature provides important diagnostic
capabilities, for instance permitting discrimination of spurious
signals due to cross-talk between the antenna elements.  Since the
antenna pattern repeats with every $120\degr$ of rotation, any signal
in the far-field remains unchanged by such a rotation; signals due to
cross-talk, however, will rotate with the faceplate.  Moreover, at the
South Pole, celestial sources track at fixed elevation, and the
ability to simulate parallactic angle rotation provides a powerful
consistency check for observations of calibrator sources, as described
in \S\ref{sec:calibrators}.  For purposes of imaging, the rotation
also allows dense sampling of the Fourier plane (see
\S\ref{sec:interferometer}).

\subsection{Primary Antenna Elements}
Each antenna consists of a 20-cm aperture, $30\degr$ semi-flare angle
corrugated horn with a lens to correct the phase front and achieve a
diffraction-limited beam on the sky.  To make the array maximally
compact, the receiver dewars were designed to fit entirely within the
footprint of the horns; the shortest baseline is $25.1$~cm.  Unlike
Cassegrain elements, horns provide unobstructed apertures, with lower
sidelobe response and better cross-talk characteristics, important for
a compact array in which the elements are nearly touching.  Each
antenna element is further surrounded by a corrugated shroud, yielding
a measured monochromatic crosstalk level of less than $-100~{\rm dB}$
in laboratory measurements of the horn-lens combination
\markcite{halverson01b}(Halverson \& Carlstrom 2001).  To further suppress cross-talk from correlated
amplifier noise, the receivers are also equipped with front-end
isolators.

Each element is equipped with a high density polyethylene lens,
resulting in a high aperture efficiency, $83.5\%$, and permitting an
extremely compact horn design.  At polar temperatures, the lens
contributes less than $2.5~{\rm K}$ to the system temperature.

The beam pattern, which determines the field of view of the
interferometer, has been characterized in range measurements at 26,
30 and 36~GHz and in each case found to agree closely with the
predicted pattern, both in the sidelobe response, which is typically
$-20~{\rm dB}$ at the first sidelobe, and in the main beam width,
which is $3\fdg4$ FWHM at 30~GHz \markcite{halverson01b}(Halverson \& Carlstrom 2001).  From the
agreement between these measurements and the theoretical beam pattern,
we adopt the theoretical value as our estimate of the true aperture
efficiency and assign a fractional uncertainty of $4\%$, as shown in
Table 1.

\subsection{Signal Chain}
Each receiver employs a cryogenically cooled, 4-stage InP HEMT
amplifier operating from 26--36~GHz.  These amplifiers were
constructed at the University of Chicago, after a design developed at
NRAO \markcite{pospiezalski95}(Pospiezalski {et~al.} 1995).  Receiver temperatures range from
15--26~K at the center of the band, increasing to an average of
$30~{\rm K}$ at the band edges.  These noise temperatures include the
HEMTs, isolator and polarizer (all at 10~K), and warm throat, horn
and lens.  Including CMB and atmosphere, we achieve typical system
temperatures of 26~K at band center, for an rms sensitivity per
visibility of approximately $60~{\rm Jy~s^{1/2}}$ in a 1 GHz band.

The 26--36~GHz RF signal from each antenna is mixed down to
2--12~GHz IF band using a local oscillator (LO) tuned to $38$~GHz.
The IF signal is split into ten 1-GHz wide bands, each of which is
further mixed down to 1--2~GHz.  The 13 signals at each frequency are
fed to one of 10 identical analog correlators \markcite{padin00}({Padin} {et~al.} 2001), where
the 78 complex multiplications are formed, digitized and integrated
for 0.84~s in a digital accumulator.  A copy of each input signal is
phase-shifted by $90\degr$, and multiplications are performed
simultaneously for the real and imaginary Fourier components.  A
$180\degr$ phase switch is applied to each LO in a Walsh sequence on a
$25.6~{\rm \mu s}$ clock interval and is demodulated by the
accumulators, to remove any offsets or slowly varying pickup.  To
further reduce residual offsets, a second, slower level of $180\degr$
Walsh switching is applied to the LOs, and is demodulated in
software with a switching period equal to the readout
interval.  The multiplier gains and quadrature errors are periodically
calibrated by injection of a correlated broadband noise source at the
input to each receiver.

Each analog correlator is integrated onto a single full-depth VME
card, and the entire 10~GHz correlator fits into a crate approximately
75~cm on a side.  Filtering and down-conversion of the IF signal is
accomplished in a similar crate, and both rotate with the antennas on
the underside of the telescope faceplate.  The short fixed distance
from the receivers to the downconverter and correlator eliminates
flexure of cables carrying phase-sensitive signals, resulting in
excellent phase stability; observed instrumental phase drifts are less
than $10\degr$ over a period of many weeks.

\section{Interferometer Characteristics}
\label{sec:interferometer}

An interferometer directly measures components of the Fourier
transform of the sky brightness.
The response on the sky for a two-element interferometer is an
interference fringe pattern, with the sinusoidal variation oriented
parallel to the baseline vector {\bf b} connecting the two elements.
The spacing of the fringe pattern is
${\lambda/b}$, where $\lambda$ is the observing wavelength and
$b$ is the magnitude of the projection of {\bf b} perpendicular
to the line of sight.  Each baseline of a multi-element array
therefore probes a point $\mathbf{u}$ in Fourier space whose
coordinates are given by $(u,v) = (b_x/\lambda, b_y/\lambda)$, or in
terms of multipole moment, $l \simeq 2\pi b/\lambda$ (for $l \gtrsim
60$) \markcite{white99a}(White {et~al.} 1999); shorter baselines measure larger angular
scales, and vice versa.

The fringe pattern on the sky is enveloped by the primary beam of the
array elements, $A(\mathbf{x},\lambda)$, where $\mathbf{x}$ is a
direction on the sky.  The output of the interferometer, the {\it
visibility}, is the time-averaged integral of this pattern, multiplied
by the sky brightness, and is the fundamental data quantity discussed
in these papers.  For observations of the CMB in the Rayleigh-Jeans
limit, the visibility in the flat-sky approximation is given by
\begin{equation}
V(\mathbf{u}) =  \frac{2k_BT}{\lambda^2}\int^\infty_{-\infty}{d{\mathbf{x}\,A(\mathbf{x},\lambda)\frac{{\Delta T}}{T}(\mathbf{x})e^{-2\pi i\mathbf{u\cdot x}}}}
\label{eqn:vis}.
\end{equation}
Since the CMB power spectrum is directly related to the Fourier
transform of the brightness $\widetilde{\Delta T}(\mathbf{u})$,
\begin{equation}
C_l\Big|_{l = 2 \pi |\mathbf{u}|} \simeq \left<\left|\frac{\widetilde{\Delta T}}{T}(\mathbf{u})\right|^2\right>,
\end{equation}
appropriate for $|\mathbf{u}| \gtrsim 10$ (\markcite{white99a}White {et~al.} 1999), a simple estimator for
the $C_l$ can be constructed from the variance of the visibilities
(see {\paperh} for a more formal discussion).
\begin{figure*}
\plotone{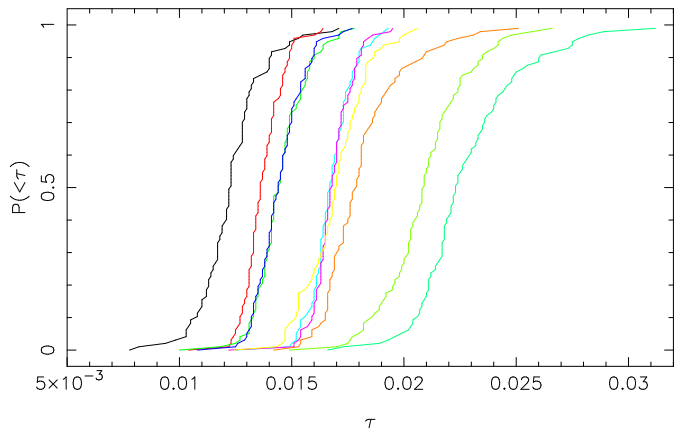}
\caption{Cumulative opacity distributions, by frequency, at the South
Pole, from 05 May--07 November 2000.  The ten curves in this plot,
from left to right, are for frequencies from 26.5 to 35.5 GHz in 1 GHz steps.}
\label{fig:tau}
\end{figure*}

With 13 elements, the DASI array yields 78 distinct baseline vectors;
the frequency dependence of the $(u,v)$ coordinates means that with 10
frequency channels, DASI samples 780 different points in the Fourier
plane (see Figure~\ref{fig:int}), although visibilities from
neighboring frequencies of the same baseline are highly correlated.
The 3-fold symmetry of the faceplate reduces the number of unique
baseline lengths to 26, each pointing of the telescope producing 3
independent measurements of the same multipole at each frequency.
With baseline separations ranging from 25.1--120.7~cm, DASI samples
multipole moments from $l=140$ at the lowest frequency to $l=900$ at
the highest.  Array parameters are summarized in Table 1.

Images of the sky can be produced from an interferometer by simple
Fourier inversion of the visibilities.  The transform of the $(u,v)$
sampling defines the effective resolution of the image, the so-called
{\it synthesized beam}; for DASI, the instantaneous $(u,v)$ coverage
produces a synthesized beam width of approximately $20^\prime$ (FWHM),
with radial sidelobes due to the radial structure of the
$(u,v)$ sampling (top right of Figure~\ref{fig:int}).

Rotation of the faceplate preserves the baseline lengths, but rotates
the fringe patterns on the sky.  This feature can be used to obtain
additional independent samples of the same multipole moment in a given
field, or for purposes of imaging, to reduce sampling artifacts in the
synthesized beam by filling in the $(u,v)$ plane.  With the 3-fold
redundancy, the pattern of $(u,v)$ coverage repeats with each $60\degr$
of rotation, allowing the same Fourier component to be measured by
independent baselines.

{
\vspace{10pt}
\footnotesize
\begin{center}
\begin{tabular}{lc}
\multicolumn{2}{c}{TABLE 1}\\
\multicolumn{2}{c}{\sc Summary of Array Parameters}\\[2pt]\hline\hline
Primary antenna elements        &$13$\\
Aperture Diameter               &$20~{\rm cm}$\\
Beamwidth (30 GHz)              &$3\fdg4\pm0\fdg07$\\
Aperture Efficiency             &$0.835\pm0.033$\\ 
Gain                            &$9.5\pm0.4~{\rm\mu K/Jy}$\\
Band 	                        &26--36~${\rm GHz}$\\
Correlator IFs                  &$10\times1~{\rm GHz}$\\
Rms Sensitivity (1 GHz band)    &$\sim60~{\rm Jy\,s^{1/2}}$\\
$B_{\rm min}$                   &$25.1~{\rm cm}$\\
$B_{\rm max}$                   &$120.73~{\rm cm}$\\[2pt]\hline
\end{tabular}
\end{center}
\vspace{0pt}
}

\section{The South Pole Environment}
\label{sec:site}

Located high on the Antarctic plateau at an altitude of 2.8~km, the
South Pole station offers a unique astronomical environment, ideal for
HEMT-amplified CMB experiments where long integration times are
required to detect signals millions of times weaker than the typical
system temperature.  Situated less than a kilometer from the
geographic South Pole, DASI can track fields continuously, with no
variation in airmass.  Sunrise and sunset at the Pole occur once per
annum, at the equinoxes; for six months of the year the sun is below
the horizon.  Rising and setting of the moon occur once every month,
yielding two weeks per month of continuous dark time.  (Naturally,
these sources never rise higher than $\sim23\fdg5$ above the horizon.)
The slow variation in astronomical conditions at the Pole contributes
to the extreme stability of the local environment; the peak-to-peak
diurnal temperature change, averaged over 1999--2001, is less than
$6\degr$C.

The Antarctic plateau is one of the driest deserts on the planet, with
annual precipitation averaging less than 8~g/cm$^2$.  Moreover,
cooling over the poles flattens the troposphere, so that the polar
atmosphere has a lower water column density than other sites of the
same physical altitude; radiosonde measurements above the Pole
indicate that the tropopause occurs between 8 and 9 km, compared to
11--13~km nearer the equator.  The precipitable water vapor column
above the pole varies between $0.25$ and $0.7~{\rm mm}$ annually
\markcite{chamberlin01}(Chamberlin 2001). 

The polar airmass is also very stable, with
surface winds dominated by a weak katabatic flow from 
gently sloping higher terrain several hundred kilometers away
\markcite{king97}(see discussion in King \& Turner 1997). 
Measured fluctuations in atmospheric 
emissivity at microwave frequencies are minimal, making the South Pole
a premier site for degree scale imaging of the CMB \markcite{lay98}({Lay} \& {Halverson} 2000). 
Our data confirm the superiority of the site---less than 5\% of
the data from the first season of observations is rejected due to 
weather (see \S\ref{sec:reduc}).

Opacities were measured with DASI from skydips performed daily during
May--November 2000.  The mean opacity determined from these data rises
from $\tau = 0.012$ to $\tau = 0.023$ over the DASI frequency band
(26--36~GHz), with little day-to-day variation; at the lowest
frequency, $95\%$ of the measured opacities are $< 0.015$, while at
the highest frequency, $95\%$ are $< 0.028$ (see Fig~\ref{fig:tau}).
At typical ambient temperatures during the winter ($-60\degr$~C) these
results indicate that over much of the DASI band, the CMB and the
atmosphere contribute roughly equal amounts to our system
temperatures.

\section{Foregrounds}
\label{sec:foregrounds}

\subsection{Diffuse Foregrounds}
\label{sec:diffuse}

Any experiment intended to measure temperature fluctuations in the CMB
must contend with a variety of diffuse Galactic foregrounds.  Chief
among these are synchrotron emission from the coupling of relativistic
electrons to the Galactic magnetic field, thermal-brehmstrahlung
(free-free) emission from ionized plasmas, and various emission
mechanisms associated with Galactic dust.

All-sky maps at frequencies below $\sim 10~$GHz will be dominated by
synchrotron emission, since the flux density scales with frequency as
$\nu^{-0.8}$ \markcite{platania98}({Platania} {et~al.} 1998), with evidence for spectral steepening
to $\sim\nu^{-1.0}$ at frequencies above 1--2~GHz \markcite{banday91}({Banday} \& {Wolfendale} 1991).
If we assume that all the emission in the maps of \markcite{haslam81}{Haslam} {et~al.} (1981) at
408 MHz is due to synchrotron, application of the more conservative of
these scaling laws implies that synchrotron can contribute at most a
few percent of the CMB anisotropy signal over DASI's $l$-range
\markcite{tegmark_ehd99}(Tegmark {et~al.} 2000).  The DASI fields are moreover selected to lie
at high Galactic latitude, corresponding to the minimum of the 408 GHz
maps.

Although the frequency dependence of free-free emission from Galactic
plasma ($\sim\nu^{-0.15}$) makes it potentially the most serious and
the most difficult to constrain of the diffuse contaminants, its
contribution to the anisotropy signal can be estimated by scaling from
$H\alpha$ maps, assuming an average electron temperature of $10^4~$K
\markcite{kulkarni87}(see, e.g., {Kulkarni} \& {Heiles} 1988).  Maps of the southern sky in $H\alpha$
\markcite{gaustad00,mccullough01}({Gaustad} {et~al.} 2000; {McCullough} 2001) indicate that the signal from free-free
should make a negligible contribution to the DASI power spectrum.

\begin{figure*}[t]
\plotone{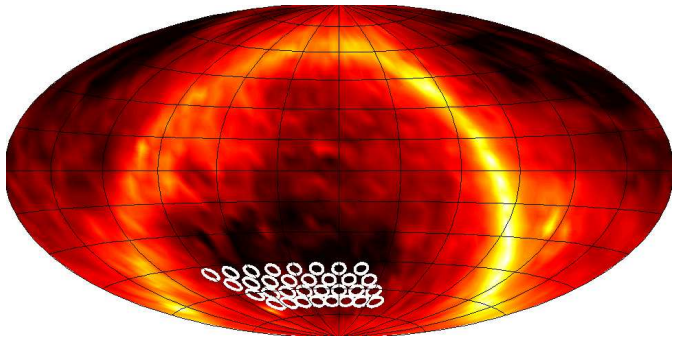}
\caption{Locations of the DASI CMB fields, plotted over the IRAS
$100~$micron map, in equatorial coordinates.  (Color map is logarithmic,
spanning 4 decades of intensity.)}
\label{fig:iras}
\end{figure*}

While thermal emission from dust will be insignificant at $\sim30~$GHz
for any reasonable range of dust emissivities and temperatures
\markcite{finkbeiner99}({Finkbeiner}, {Davis}, \&  {Schlegel} 1999), observations reported by the OVRO Ring
experiment \markcite{leitch97}(Leitch {et~al.} 1997) suggest that non-thermal emission
associated with the dust may contaminate anisotropy measurements at a
level comparable to the CMB signal in the $l\sim600$ range, at least
at frequencies near 15~GHz.  Although little is known about the
amplitude or homogeneity of this component, the strong spatial
correlation with IRAS 100 micron maps indicates that it is well traced
by the thermal dust emission; field positions are therefore chosen to
coincide with the minimum in the IRAS 100~micron map of the southern
sky (see Figure~\ref{fig:iras}), where typical dust intensities are a
factor of $\sim 5$ lower than in the OVRO fields.

While the current state of our knowledge of all of these Galactic
foregrounds implies that the CMB will dominate the anisotropy signal
at the frequencies and angular scales to which DASI is sensitive
\markcite{tegmark_ehd99}(see, e.g., Tegmark {et~al.} 2000,  for a comprehensive review), we
employ an additional check which is independent of the assumptions and
scaling laws presented above.  As discussed in {\paperh}, when
extracting the CMB power spectrum from the visibility data, we project
out foregrounds on the basis of their spatial template alone, with no
assumptions about the intensity of the various components at
centimeter wavelengths.  These constraints restrict the contribution
of any of these diffuse foregrounds to be less than $\sim 3\%$ of the
CMB signal, assuming only that the 26--36~GHz counterparts to these
foregrounds have the same spatial distribution as they do at
frequencies where the foregrounds are dominant; it is assumed that
each diffuse foreground can be described with a constant single spectral
index in a given DASI field.

\subsection{Discrete Sources}

As expected for cm-wave, small angular scale experiments, point
sources are the dominant astronomical foreground for DASI \markcite{tegmark96}(see,
e.g., Tegmark \& Efstathiou 1996), and the dominant contributor to the measured
anisotropy.  The method described above for projecting out diffuse
foregrounds can be applied to point sources as well, since they
produce a unique spatial signature in the visibility data.  To null
the effect of point sources through this technique, we therefore
require only the positions of the sources, {\it not their fluxes}.

Positions for the brightest point sources are determined from the DASI
data themselves; the procedure for identifying bright point sources is
described in detail in \S\ref{sec:ptsrc} below.  Positions for point
sources below the DASI detection threshold are determined from the PMN
catalog \markcite{wright94}({Wright} {et~al.} 1994); in all, we constrain every point source from
the PMN survey whose 5 GHz flux density, when modulated by the DASI
primary beam, exceeds 50 mJy.

For sources below the PMN flux cutoff, we require a statistical
correction to the power spectrum.  These corrections, given explicitly
in {\paperh}, are determined by Monte Carlo simulation of a point
source population with $dN/dS$ given by the PMN catalog, and whose
spectral index distribution is determined by observation of a complete
sample of PMN sources with the OVRO 40-meter telescope at 26--36~GHz
(paper in preparation).

\subsection{Ground Contamination}
\label{sec:ground}

Astronomical foregrounds aside, the presence of ground contamination
at a level much greater than the expected cosmological signal places
the most stringent constraint on our observing strategy.  Although DASI
was intended to operate with a ground shield, the panels could not be
installed until the 2000--2001 austral summer and were not in place
while the data described here were collected.

The signature of the ground is visible even in short integrations; as
fields are tracked over the full azimuth range, excluding the MAPO
building, variations as large as $7\,\sigma$ above the expected
thermal noise ($\sigma\sim 4$~Jy) can be seen in 5-minute integrations,
while directly over the roof of MAPO, 20-$\sigma$ signals are seen on
a few of the shortest baselines.  However, these fluctuations are
strongly dependent on baseline length, falling sharply with increasing
$(u,v)$ radius; on the longest baselines, the raw visibilities are
consistent with thermal noise at all frequencies, even when observing
directly over the MAPO building.

Although the ground shows a strong azimuthal dependence, repeated
24-hr tracks demonstrate that the signals are quite stable over time.
When data taken as much as five days apart are differenced, the
residuals are consistent with the thermal noise, even on the shortest
baselines, and over the full azimuth range, including directions
towards MAPO and other station buildings.  

\section{Observations}
\label{sec:obs}

\subsection{Field Selection}
\label{sec:fld}

Because the ground signal exhibits long-term stability, our strategy
was to observe independent fields over a narrow azimuth range on
timescales short enough that the ground could be expected to remain
constant. Observations were divided among groups of 8 fields separated
by 1\,h of RA, each group at a constant elevation.  Each field in a
group was observed over a fixed 1-hr azimuth range; over the course
of 8 hours, these observations yield 8 samples of the CMB sky, and 8
independent samples of the same ground, allowing a constraint on the
common mode (and reducing the effective number of degrees of freedom
available for constraining the CMB power spectrum).  Observations were
timed so that data from each field sampled precisely the same ground.
Azimuth ranges were selected which avoided
lines of sight passing over any of the South Pole station or science
buildings, or the communications antenna field.

Four such rows of 8 fields, referred to as the A--D rows, were
arranged on a hexagonal grid (to facilitate eventual mosaicing of the
fields), spaced by 1\,h in RA and $6\degr$ in declination. The grid of
fields was positioned at high elevation ($\ge49\degr$), to further
reduce the signal from the ground.  Besides permitting a constraint on
ground contamination, as discussed above, the spacing of 1\,h in RA
results in negligible inter-field correlations, greatly simplifying
the analysis described in {\paperh}.  Declinations and right
ascensions of the first field in each row are given in Table~2.

{
\vspace{10pt}
\footnotesize
\begin{center}
\begin{minipage}[t]{3in}
\begin{tabular}{lllc}
\multicolumn{4}{c}{TABLE 2}\\
\multicolumn{4}{c}{\sc CMB Field Row Positions\footnote{\rm Field
positions are obtained by adding 0--7\,h to the RA listed for each row.}}\\[2pt]\hline\hline
Row & RA (J2000)           & Dec (J2000)         & Days Observed \\\hline
A   & $22\,\,00\,\,00$  & $-61\,\,00\,\,00$ & 14\\
B   & $21\,\,30\,\,00$  & $-67\,\,00\,\,00$ & 24\\
C   & $22\,\,30\,\,00$  & $-55\,\,00\,\,00$ & 28\\
D   & $23\,\,00\,\,00$  & $-49\,\,00\,\,00$ & 31\\\hline
\end{tabular}
\vspace{-10pt}
\end{minipage}
\end{center}
\vspace{10pt}
}

\subsection{Observing Schedule}

As described above, in each 24-hr period, we obtained a total of 16
hours of integration on 8 CMB fields: 8 hours over each of two azimuth
ranges.  The remainder of a 24-hr period was divided among various
calibration tasks.  Every 8-hr period was bracketed by 1.5-hr scans
on a calibrator source, as discussed in the next section, and these
11-hr periods were in turn bracketed by skydips, yielding two
independent measures of the atmospheric opacity daily.  These data are
described in \S\ref{sec:site}.  Finally, at the beginning and end of
every 24-hr period, the noise source was used to calibrate the
complex correlator, as described in the next section.

\section{Calibration}
\label{sec:calibration}

Several kinds of calibrations are performed for each of DASI's 78
baselines, in each of the 10 bands: 1) calibration of the complex
multipliers to correct for non-orthonormality of the real and imaginary
visibility response, 2) absolute calibration of the flux scale for
each baseline, which is transferred to a celestial source and 3)
regular observations of celestial calibrators, to refer visibility
amplitudes to the absolute flux scale and visibility phases to a phase
center.

The second and third of these calibration tasks are described in
separate sections below.  The first, the calibration of the complex
multipliers, was accomplished by injection of a strong signal from the
correlated noise source.  This calibration was performed at the
beginning and end of every 24-hr period, although for $95\%$ of all
baselines, the rms variation of the relative gain and phase offset
between real and imaginary multipliers were less than $2\%$ and
$1\fdg2$, respectively, over 97 days.

\subsection{Absolute Calibration}
\label{sec:abscalibration}

The relative scarcity of sources whose flux densities are known at
high frequencies makes absolute calibration of any microwave
experiment a challenging proposition, as does the generic steepness of
most radio source spectra.  Additionally, of the sources which are
well studied, few are accessible from the South Pole.  Planets never
rise more than $\sim23\fdg 5$ above the horizon, and the best studied
of these, Jupiter and Mars, were either very low or below the horizon
throughout the 2000 observing season.  Moreover, with DASI's 20-cm
apertures, beam dilution makes it inherently difficult to detect all
but the brightest point sources; with a gain of approximately $10~{\rm
\mu K/Jy}$, to achieve a S/N $\sim1$ per visibility on a 1 Jy source
requires about an hour of observation.

Consequently, the absolute calibration of DASI is based on
measurements of beam-filling external thermal loads of measured
temperature.  The noise source power level (which is separately
monitored for drift) was calibrated for each receiver from these
measurements, and a calibration of correlated noise source power was
derived for each baseline.  This calibration was immediately
transferred from the noise source to a celestial source.  

For the transfer observations, the source was tracked for 3 hours at
each of six faceplate rotations, separated by $60\degr$, with a
reference field observed for ground subtraction.  Due to the
three-fold symmetry of the antenna locations, this procedure yielded
six independent measurements of each $(u,v)$ point from different
baselines with different contributions from the ground.  The source
model was averaged over these observations to reduce any residual
systematics, resulting in a determination of the flux for each
baseline with a statistical uncertainty of $<2\%$.  

The load measurement and transfer procedure was conducted once during
the austral summer of 2000 and again in the summer of 2001.  The
calibration of 2001, on which the absolute scale of the DASI
observations is based, was transferred directly to {\RCW}.  The 2000
calibration was transferred without ground subtraction to {\carneb}
and later to {\RCW} through interlaced observations of the two
sources.  We estimate the statistical uncertainty on the overall flux
scale resulting from the load measurement procedure and transfer
observations to be $<1\%$, averaged over baselines.  Consistent with
this estimate, the average ratio of fluxes derived from the
independent calibrations of 2001 and 2000 is $0.997$.  The systematic
uncertainty common to these two calibrations, resulting mainly from
uncertainty in the determination of antenna coupling to the loads and
in the effective load temperatures, we estimate to be $3\%$, and is
the dominant contribution to the total $3.5\%$ uncertainty in the
overall flux scale.

\subsection{Celestial Calibrators}
\label{sec:calibrators}

As previously noted, each 8-hr period of CMB observations was
bracketed by 1.5-hr scans of celestial calibrators to refer the
visibility amplitudes to the absolute flux scale and to refer the
phases to a common phase center.  In addition, the correlated noise
source was injected every hour to track short-term system stability.
In practice, the instrumental phase and amplitude response were quite
stable; over 97 days of observation, the phases showed a typical rms
variation of less than $15\degr$.

For the fields referred to as the A row in Table~2, the amplitude
calibrator used was {\carneb}, a well-known Galactic HII complex.  The
Carina Nebula's free-free spectrum makes it a good high frequency
calibrator; the source is bright enough that its flux can be measured
to a few percent on most baselines in approximately 15 minutes.
However, the source is dominated by an extended central region, with
much weaker flux on smaller scales, necessitating significantly longer
integrations on the longest baselines to achieve uniform accuracy.

These considerations led us to use {\RCW}, a more compact Galactic HII
region, as the flux calibrator for the B--D row observations.
Although its integrated flux is lower than that of {\carneb}
\markcite{leitchiau}(see {Leitch} {et~al.} 2000), the source is readily detectable, with
nearly uniform flux ($\sim 150~{\rm Jy}$) out to the longest
baselines; for $(u,v)$ radii $> 80~\lambda$, the source is
considerably stronger than Carina.

Although {\RCW} is partially resolved by the longest baselines,
observations of the source while rotating the faceplate show that the
rms phase variation with parallactic angle, averaged over all
baselines, is less than $10\degr$.  These observations indicate that
{\RCW} is sufficiently point-like (or at least radially symmetric)
that the source can be used as a phase calibrator.  Although {\RCW} is
located in the Galactic plane, these observations also demonstrate
that any contaminating flux is small.

Thus, for the B--D row data, each 1.5-hr scan was spent observing
{\RCW}, which was used as both amplitude and phase calibrator.  For
the A-row observations, this time was split between {\carneb} and the
extragalactic source Centaurus A, which served as the phase
calibrator.  On all but the longest baselines of the A-row data, these
observations yielded measurements of both phase and amplitude to
better than $2\%$.

\section{Data Reduction}
\label{sec:reduc}

The data presented in this paper (and analyzed in Papers II and III),
comprise 97 days of observations obtained during 05 May--07 November
2000, divided among 32 fields, as described in \S\ref{sec:fld}.  These
data represent an observing efficiency of approximately $85\%$ of the
time devoted exclusively to CMB observations; the remainder of the
time was lost to hardware maintenance and reliability.  Because our
ability to constrain the ground depends critically on uniform sampling
of a fixed azimuth range for each of 8 fields, as described in
\S\ref{sec:ground}, a draconian pre-edit is applied to the CMB data --
the loss of all data for a single field for any reason results in the
rejection of the entire 8-hour period for that baseline and correlator
band.

Visibilities are accumulated from each correlator in 8.4-s
integrations, along with monitor data from the telescope drive systems
and receivers.  Prior to combination for input to the power spectrum
analysis described in {\paperh}, various edits are applied to these
data, falling into three general categories: cuts on monitor data,
cuts derived from the visibility data themselves, and calibration
edits.

In the first category, data are rejected when a receiver is not
operating, when a receiver LO loses lock, or when either the 10 or
50~K cryogenic receiver stages shows a gross warming trend; these cuts
typically reject $\sim5\%$ of the data.  Next, data are edited on the
quadrature errors in the complex multipliers, determined twice daily
by injection of a broad-band noise source.  Baselines for which the
relative gain between the real and imaginary multipliers falls outside
the range $0.9\pm0.3$ (deviation of the mean value from 1 reflects the
signal path asymmetry introduced by the hybrid splitters), or for
which the phase offset falls outside the range $0\degr\pm20\degr$, are
rejected.  These ranges are determined from the empirical distribution
of quadrature errors and reflect the values at which the distributions
become significantly non-Gaussian; these data are interpreted as
evidence for microstrip termination errors or malfunctioning
multiplier chips and showed little variation throughout the 2000
season.  When the raw visibilities are calibrated for these offsets,
data are rejected if the bracketing values differ by more than $10\%$.
Cuts on the quadrature errors collectively reject $\sim11\%$ of the
data.

The closest approach of the moon to the DASI fields during their
observation is $36^\circ$. Fringing from the moon is evident on the
shortest baselines, and for $(u,v)$ radii $<40\,\lambda$, we reject
all data when the moon is above the horizon.  The D-row observations
were made from 11 September--11 October, bracketing the sunrise, with
an additional week obtained in early November; data for the D-row
fields are also excluded for $(u,v)$ radii $<40\,\lambda$ when the sun
is above the horizon.

In the second category of edits, the raw visibilities are combined
into 1-hr bins, and the baseline-baseline correlation matrix is
computed separately for each correlator.  Large off-diagonal elements
of the correlation matrix are interpreted as evidence for
contamination by atmospheric fluctuations.  By comparison of data for
the same field from different azimuth ranges on the same day, or from
different days, we find that the precise threshold of the correlation
cut does not strongly affect data consistency, and we reject an 8-hr
observation for which any off-diagonal element exceeds $\pm0.36$,
affecting $5\%$ of the data.

In the last category, data are edited on the quality of the associated
observations of celestial phase and amplitude calibrators.  Data are
rejected when the bracketing calibrator amplitudes vary by $> 10\%$, or
when the bracketing phases change by $> 30\degr$.  Data are also
rejected when previous edits have reduced the sensitivity on a
calibrator scan by $> \sqrt{2}$, resulting in a statistical error of
$> 3\%$.  On average, these cuts reject $\sim20\%$ of the data. Overall, the
combined edits from all categories reject a total of 40\% of the data.

For input to the power spectrum analysis described in {\paperh}, the
edited and calibrated data are combined into 1-hr integrations,
yielding a maximum of 1560 visibilities per field (78 complex
baselines$\times$10 correlator channels) for each observation.

\section{Point Sources}
\label{sec:ptsrc}

The brightest point sources in our fields are obvious in the
synthesized maps (see Figure~\ref{fig:int}), even without ground
subtraction.  However, as discussed in {\paperh}, and in
\S\ref{sec:foregrounds} above, to extract the CMB power spectrum from
the visibility data, we employ a method capable of projecting out
point sources on the basis of their positions alone.  As described in
{\paperh}, we construct the list of point sources to null using a
combination of source positions derived from the DASI data, as well as
positions of sources in the PMN southern catalog.  Here we describe
the procedure used to find sources in the DASI data.

\begin{figure*}[t]
\plotone{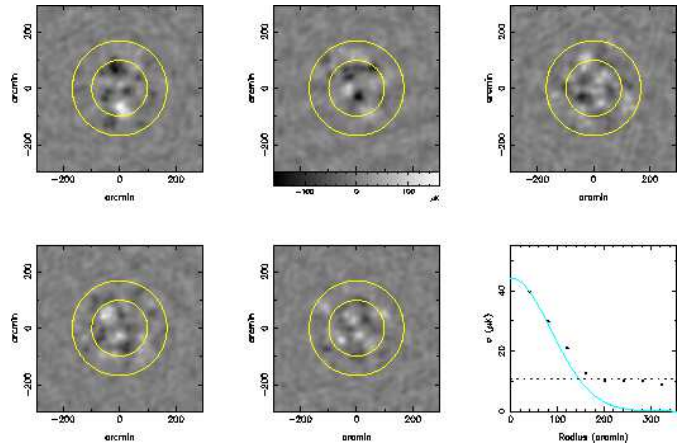}
\caption{Images of five DASI CMB fields. The two
concentric circles represent the $-3$~dB and $-10$~dB taper of the
beam, respectively. The lower right panel shows the rms pixel values
for field 5 (lower middle panel) as a function of radius (black
points), the primary beam taper normalized to the first rms pixel
value (solid line), and the theoretical rms image noise, determined
from the scatter in 8.4-s visibility data (dashed line).}
\label{fig:flds}
\end{figure*}

For each set of fields, images are made from ground-subtracted
visibilities using the DIFMAP imaging package \markcite{shepherd94}({Shepherd}, {Pearson}, \&  {Taylor} 1994).
When making the images, baselines $< 64~\lambda$ are excluded; this
cut removes the bulk of the CMB power in the data, significantly
increasing the S/N for point source detection.  The maps are then
divided into annular bins around the field center; the pixel rms in
each bin, averaged over 8 fields, is fit to a noise model, consisting
of a signal with the shape of the primary beam, added to a flat
detector noise floor (as in the bottom right panel of
Figure~\ref{fig:flds}).  Each image is then divided by the fitted
noise model to produce an effective S/N map.  The highest peaks,
multiplied by a copy of the synthesized beam, are then iteratively
subtracted from the image until no peaks $> 4.5$ remain; we estimate
that in the absence of point sources, 0.03 peaks above this threshold
should occur in each set of 8 fields.

Using this technique, we can detect a 40~mJy source at beam center
with $>4.5$-$\sigma$ significance.  In the 32 DASI fields, we find 28
sources, whose estimated $S_{31}$ fluxes range from 80~mJy to 7~Jy.
Correlating these locations with the PMN catalog, we find
$S_5>100$~mJy counterparts for them all, with probabilities for
accidental association $\lesssim0.01$.  Positional errors are
typically several arcminutes.  However, note that the PMN survey will
resolve some of the closest sources, yielding positions which may
differ from the effective position as seen by DASI.  This limits the
determination of our absolute pointing error, but we estimate it to be
better than $2^\prime$.

Fitting the position of the very brightest sources for each day of
data over the observations of a given set of fields shows that
pointing drift over $\sim30$ days is $<1'$.  Servo tracking jitter is
$\ll1'$ and the fact that point sources appear in the maps with
exactly the shape of the synthesized beam confirms this.

\section{Results}
\label{sec:results}

Although Fourier plane visibilities are the natural data product of an
interferometer, image-plane analysis provides a valuable consistency
check on the data quality.  Shown in Figure~\ref{fig:flds} are a
subset of the DASI fields, after point source removal and ground
subtraction.  In these images, point sources identified in the DASI
data were removed by subtraction of a delta function model from the
visibilities, while ground contamination has been removed by
subtraction of a mean visibility, averaged over 8 fields, from each of
the 1-hr field integrations.  Artifacts of the $(u,v)$ sampling have
been reduced in these maps by a variant of the CLEAN algorithm
\markcite{hogbom74}(H\"{o}gbom 1974), restricted to the $-10~{\rm dB}$ contour of the
primary beam, similar to the procedure described in \S\ref{sec:ptsrc}
for finding point sources.

For an interferometer, each visibility is convolved with the
autocorrelation of the antenna aperture; in the image plane, this
translates into an enveloping of the image by the primary beam.  Here
the images have been deliberately extended to radii well beyond the
central beam area, so that detector noise should dominate near the
edges.  As can be seen qualitatively in Figure~\ref{fig:flds}, after a
simplistic removal of the ground and point source signals, every field
shows residual structure in the central region of the beam; as
demonstrated in the lower right panel, the rms in radial bins follows
the beam profile, tapering to the theoretical noise floor, typically
$10~{\rm \mu K}$ (at $\sim20^\prime$ resolution), far from the beam
center.

Images of the same fields observed at different epochs show
repeatable structure, and when images are constructed from
visibilities divided into two frequency ranges, the ratio of pixel
values is consistent with a thermal spectrum \markcite{leitchiau}(see {Leitch} {et~al.} 2000);
rigorous, Fourier-plane analogues of these tests are described in
{\paperh}.

\begin{figure*}
\epsscale{2}
\plotone{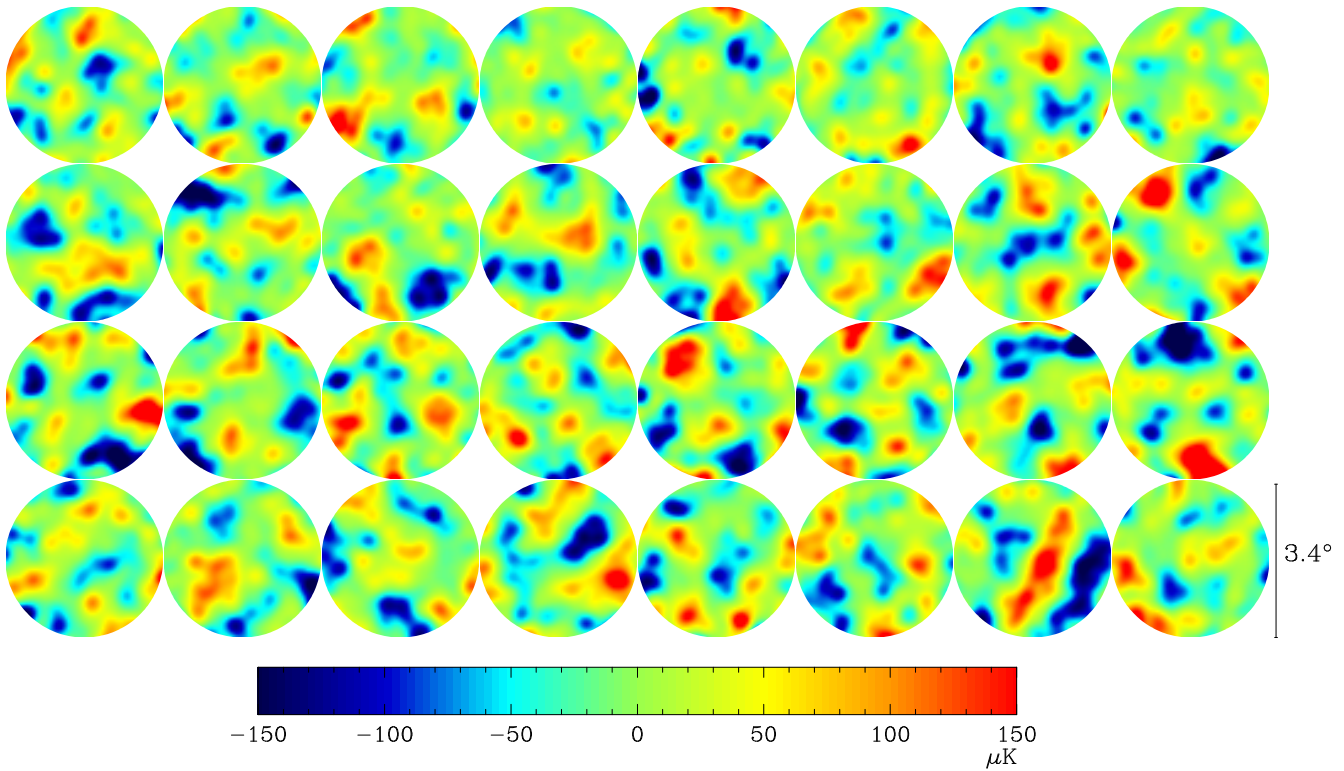}
\caption{Images of the 32 DASI fields, at $20^\prime$ resolution.
Shown are CLEANed maps of the central $3\fdg4$ FWHM of each primary
beam, corrected for the beam taper.  Typical rms noise in a
$20^\prime$ beam is $10\,{\rm\mu K}$ at map center; note however, that
because these images have been divided by the primary beam, the
effective noise increases by a factor of 2 at the map edges.  Rows are
in order of decreasing elevation: B, A, C, D from bottom to top, with
RA increasing to the left.}
\label{fig:dasiflds}
\end{figure*}

The entire complement of 32 DASI fields is shown in
Figure~\ref{fig:dasiflds}, after ground and point source removal.
Here we show only the central FWHM of each field, corrected for the
primary beam taper at the center of the band.  Although the S/N is
high enough in all these maps that the apparent structure is real,
note that the sensitivity is not uniform across the images, but
decreases by a factor of 2 at the edges due to the primary beam
correction.

We stress that these images are for presentation purposes only; no
images are used in our power spectrum analysis, nor are point sources
or ground ever subtracted from the visibility data; these foregrounds
are modeled in a self-consistent way as part of the constraint matrix
formalism presented in {\paperh}.

\section{Conclusions}

We have described the instrumentation, observations and first year
data from DASI, a novel interferometric experiment to measure
anisotropy in the CMB.  During its first season of operation, the
instrument functioned extremely well, producing 97 days of
high-quality data on 32 CMB fields, for an average of 24 hours of
integration per field, and achieving close to the theoretical noise
limits.

With these data, we are able to image $\sim400$ square degrees of sky
to a typical rms sensitivity of $10~\mu$K (at $\sim20^\prime$
resolution); we detect structure in the CMB with high S/N in every
field.  Although the ground has proven to be our largest contaminant,
the slowly varying character of this foreground and careful experiment
design enables us reliably to recover the CMB component, even in the
presence of near-field signals many times the amplitude of the
intrinsic fluctuations.  As detailed in {\paperh} and {\paperp}, these
data yield a sensitive new measurement of the CMB power spectrum and
provide important constraints on cosmological parameters.

\acknowledgments

We express our gratitude for the support provided by the Center for
Astrophysical Research in Antarctica (CARA), in particular the efforts
of Al Harper, Stephan Meyer, Fred Mrozek, Nancy Odalen, Bob Pernic,
Dave Pernic, Joe Rottman and Mark Thoma.  We thank Antarctic Support
Associates and the 1999--2000 South Pole Station construction crew for
extraordinary on-site support, and Eric Chauvin at Vertex/RSI for
professional excellence in the design and delivery of the telescope
mount. We thank the observatory staff of the Australia Telescope
Compact Array, in particular Bob Sault and Ravi Subrahmanyan, for
their generosity in providing initial point source observations of the
DASI fields.  This research is supported by the National Science
Foundation under a cooperative agreement (OPP 89-20223) with CARA, a
National Science Foundation Science and Technology Center.  Support at
Caltech is provided by NSF grants AST 94-13935 and AST 98-02989.

\bibliography{}

\end{document}